# Lusifer: LLM-based User Simulated Feedback Environment For online Recommender systems

Danial Ebrat
*School of Computer Science*
*University of Windsor*
Windsor, Ontario, Canada
ebrat@uwindsor.ca

Eli Paradalis
*School of Computer Science*
*University of Windsor*
Windsor, Ontario, Canada
pardali1@uwindsor.ca

Luis Rueda
*School of Computer Science*
*University of Windsor*
Windsor, Ontario, Canada
lrueda@uwindsor.ca

*Abstract*— Reinforcement learning (RL) recommender systems often rely on static datasets that fail to capture the fluid, ever-changing nature of user preferences in real-world scenarios. Meanwhile, generative AI techniques have emerged as powerful tools for creating synthetic data, including user profiles and behaviors. Recognizing this potential, we introduce Lusifer, an LLM-based simulation environment designed to generate dynamic, realistic user feedback for RL-based recommender training. In Lusifer, user profiles are incrementally updated at each interaction step, with Large Language Models (LLMs) providing transparent explanations of how and why preferences evolve. We focus on the MovieLens dataset, extracting only the last 40 interactions for each user, approximately 30% of their training data, to emphasize recent behavior. By processing textual metadata—such as movie overviews and tags—Lusifer creates more context-aware user states and simulates feedback on new items, including those with limited or no prior ratings. This approach reduces reliance on extensive historical data and facilitates cold-start scenario handling and adaptation to out-of-distribution cases. Our experiments compare Lusifer with traditional collaborative filtering models, revealing that while Lusifer can be comparable in predictive accuracy, it excels at capturing dynamic user responses and yielding explainable results at every step. These qualities highlight its potential as a scalable, ethically sound alternative to live user experiments, supporting iterative and user-centric evaluations of RL-based recommender strategies. Looking ahead, we envision Lusifer serving as a foundational tool for exploring generative AI-driven user simulations, enabling more adaptive and personalized recommendation pipelines under real-world constraints.

## I. INTRODUCTION

In the digital era, recommender systems are critical for personalizing content to user preferences. The evolution of these systems, particularly those using reinforcement learning (RL), introduces complex challenges in training. Recent advancements have shifted towards using simulated environments and Large Language Models (LLMs) to mimic user interactions better and integrate real-time feedback. This section briefly addresses these developments in this diverse field, highlighting the transition from traditional to LLM-based simulations and discussing their potential and limitations in training RL models.

Training environments are traditionally offline, based on historical data, which may lead to distribution shifts and limited exploration, or online, offering dynamic interactions, though struggling with sample efficiency and real-world generalization [1]. Within this context, specific simulation environments such as Recsim NG [2], Recogym [3], and Virtual-Taobao [4] have offered varied benefits and faced distinct challenges. Recsim NG's customizability and complex user models come at the cost of computational efficiency and scalability, given the various modules involved. Recogym's accessible setup for quick prototyping lacks the ability to simulate complex user behaviors, and Virtual-Taobao's realistic e-commerce simulations are limited in domain applicability.

The advent of LLMs as simulated environments introduces a promising alternative that combines flexibility, broad applicability, and the capacity to generate nuanced, realistic user feedback without previous systems' complexity or computational demands. This advancement is poised to address the limitations of existing simulation environments regarding realism, domain specificity, and scalability, potentially revolutionizing RL-based recommender systems.

Further exploration into LLM capabilities, such as User-Guided Response Optimization (UGRO) by [5], Character-LLM [6], DITTO [7], and [8]showcase attempts to optimize dialogue responses, simulate specific characters, advance role-play capabilities, and LLMs capabilities in simulating dialogues respectively. These approaches demonstrate the versatility of LLMs in simulating user interactions and behaviors, although they face challenges in achieving a high level of accuracy, lack of generalization power, and reproducible experiments.

Moreover, the LLM Interaction Simulator (LLM-InS) [9] and the exploration of Imitation of Subrational Behavior [10], both delve into simulating user-item interactions and learning subrational agent policies. These studies validate LLMs' effectiveness in addressing specific challenges like the cold start problem in collaborative filtering and modeling complex human behaviors, highlighting both the potential and the limitations associated with the quality of the underlying synthetic data being generated.

Additionally, investigations into the coordination skills of LLM-based agents in complex environments, as seen in MetaAgents [11], and the impact of persona variables on LLM performance [12] further elucidate the nuanced capabilities and challenges of LLMs in replicating human-like interaction and behavior.

Studies such as those of [13] and USimAgent [14] extend the exploration into LLMs' capabilities in simulating user behavior across various domains, emphasizing the intricacies of causal inference and the complexity inherent in accurately simulating cognitive processes such as reasoning and planning.

In this work, we present Lusifer, a generative framework used to generate users and their behaviors in the context of a recommender system. More specifically, the platform deploys a novel LLM-based simulation environment highlighting how user preferences evolve over time in recommender systems. Rather than focusing solely on improving standard accuracy metrics, Lusifer tracks and updates user profiles at each interaction step, revealing how preferences shift when encountering new item attributes or styles. These updates are accompanied by concise explanations, offering insights into why the simulated user responds differently to evolving recommendations. Figure 1 indicates the pipeline in generating the summary of the user's behavior and the way it highlights updates after each batch.

Using the movie recommendation problem as a case study, we concentrate on each user's recent interactions to assess Lusifer's performance under limited data conditions. Although Lusifer may not outperform traditional methods in standard accuracy measures, its strengths lie in configurability, adaptability, and explainability. It can handle out-of-distribution scenarios, support new item domains, and serve as a scalable, ethical alternative to live user experimentation. Furthermore, this user-centric simulation enables researchers to investigate how user profiles evolve under various recommendation strategies, providing valuable insights for training and testing RL-based recommender systems. The main contributions of this paper are the following:

(i) Dynamic, Explainable User Simulation: We introduce Lusifer as the first LLM-based environment that can generate users and behaviors to simulate evolving user feedback, storing and updating profiles at each interaction step while providing transparent explanations for changes.

(ii) Reinforcement Learning Compatibility: We demonstrate how Lusifer serves as a platform for training and evaluating RL-based recommender agents, offering a more realistic and interactive setting than static benchmarks.

(iii) Adaptability to Limited and Evolving Data: We validate Lusifer's ability to capture shifting preferences with only a partial slice of user histories, highlighting how the environment can adapt to data-scarce conditions and even out-of-distribution or cold start scenarios.

(iv) Scalable, Ethical Alternative to Live Experiments: By offering a controllable user simulation, Lusifer enables researchers to systematically explore recommendation strategies without relying on costly or potentially invasive live user trials.

## II. METHODOLOGY

In this section we discuss different stages of the methodology including data preprocessing, creating user profiles and generating simulated ratings.

### A. Data preprocessing

This study employs the MovieLens datasets [15] to evaluate Lusifer. To enrich this dataset with additional descriptive information, we utilized the TMDB API [1], a web service that provides API to a comprehensive movie information database maintained by The Movie Database (TMDb). We retrieved movie overviews, including metadata, to enhance the richness of the data and enable a more comprehensive evaluation of Lusifer.

We used different versions of the MovieLens dataset, including the 100K and 1M versions, standard benchmarks in recommender systems research [15]. Each dataset provides user ratings along with demographics and movie details. All versions of the datasets have been split into five subsets for 5-fold cross-validation, using 80% of the data for training and 20% for testing.

For users, we extracted demographic details such as age, gender, and occupation to form a comprehensive text feature called user_info, encompassing all relevant information about each user. For movies, we consolidated titles, genres, and overviews into a single text feature known as movie_info, encapsulating all pertinent information about each movie.

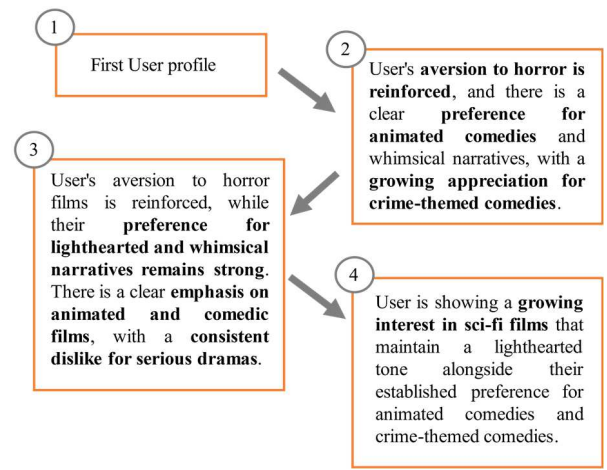

Fig. 1. Example of a user profile update and progress.

### B. Phase 1: Creating User Profiles

Our methodology unfolds in two phases: (i) creating user profiles and (ii) generating simulated ratings. In Phase 1, we constructed detailed user profiles and summarized their behavior patterns in the initial phase. The data is organized by extracting each user's last 40 rated movies from the training set while maintaining the chronological order of interactions based on timestamps. This step ensures a clear view of recent user activity. As a result, on average, we only used 30% of the training set to capture the user's behavior. We create an LLM-driven simulation environment compatible with the OpenAI API (GPT-4o-mini model) [2] and any open-sourced LLM included in Ollama [3]. For each user, the LLM was provided with user_info and historical ratings, which included movie_info for the last 40 movies rated by the user. We chose 40 to process the data in four batches of ten, yielding four distinct transitions, a balance between capturing multiple profile updates, explainable shifts in user preferences, and not complicating the simulation process. This approach also addresses the LLM's performance limitations with large input prompts. The LLM can accurately extract relevant information and update user behavior after each batch by processing historical interactions in smaller chunks. The LLM processed each batch sequentially, initially analyzing the user's characteristics and generating a summary based on the first batch. The LLM updates this summary with each subsequent batch, refining the user profile and simulating behavior changes. Additionally, Lusifer stores every intermediate user profile state. This approach enables a transparent record of how a user's preferences evolve with each new batch. Moreover, Lusifer stores concise explanations detailing the updates in user preferences after each iteration, adding an explainability layer. By tracking and validating these gradual changes, researchers can observe how Lusifer captures the dynamic nature of user behaviors across incremental interactions. The resulting summaries detailed user preferences and aversions, enabling accurate predictions of potential ratings.

---

[1] TMDB API: https://developer.themoviedb.org/reference/intro/getting-started
[2] OpenAI API: https://platform.openai.com/
[3] Ollama: https://ollama.com

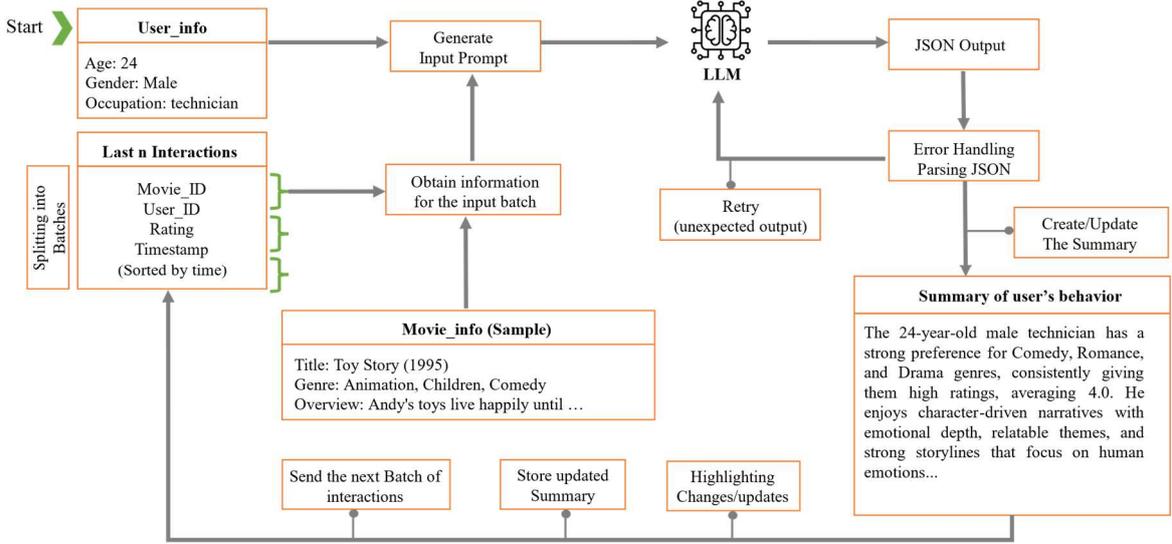

Fig. 2. Schematic view of Phase 1: creating user profiles.

### C. Phase 2: Generating Simulated Ratings

In the second phase, we generated simulated ratings based on the updated user summaries. We utilized the comprehensive overview of each user's behavior from phase 1, along with the last ten rated movies (including their information and ratings), to the LLM. Acting as the user, the LLM-rated recommended movies from the test set. The output was generated in a JSON file containing movie_IDs and corresponding ratings. We then evaluated these LLM-generated ratings against the actual ratings in the test set using Root Mean Squared Error (RMSE), Means Absolute Error (MAE), and Pearson correlation metric to evaluate the accuracy of Lusifer's performance.

### D. Baseline Methods for Comparative Analysis

To assess the performance of Lusifer in terms of predictive accuracy, we implemented several established recommendation algorithms as baselines: Alternative Least Square (ALS) [16], Singular Value Decomposition (SVD++) [16], Neural collaborative filtering (NCF)[17], and RNN4REC [18]. These methods are well-regarded in the recommender systems field for their ability to effectively model user preferences and item relationships, even compared to state-of-the-art methods, as comprehensively expressed in [19].

For consistency and to gain additional insights, we limited all baseline methods using only the last 40 interactions for each user, aligning with Lusifer's usage of the training sets. We standardized training across baselines with 50, using an embedding dimension of 16 for all latent factor models. ALS was implemented with L2 regularization ($\lambda=5.0$) and confidence weighting ($\alpha=10$) to handle implicit feedback. The SVD++ configuration incorporated user-specific biases and implicit interactions through its extended factorization formulation. For NCF, we designed a multi-layer perceptron pathway with hidden unit dimensions (128, 64, 32) and negative sampling to optimize pairwise ranking. The RNN4Rec architecture employed LSTM layers with hidden state sizes (16, 16), processing sequences of recent_num=10 interactions, complemented by negative sampling for contrastive learning. These parameter choices aimed to balance model capacity with computational efficiency while maintaining comparability through shared design constraints. By maintaining consistent data usage and specifying key parameters for these well-established algorithms, we aimed to provide a fair and insightful comparison between Lusifer and conventional recommendation techniques.

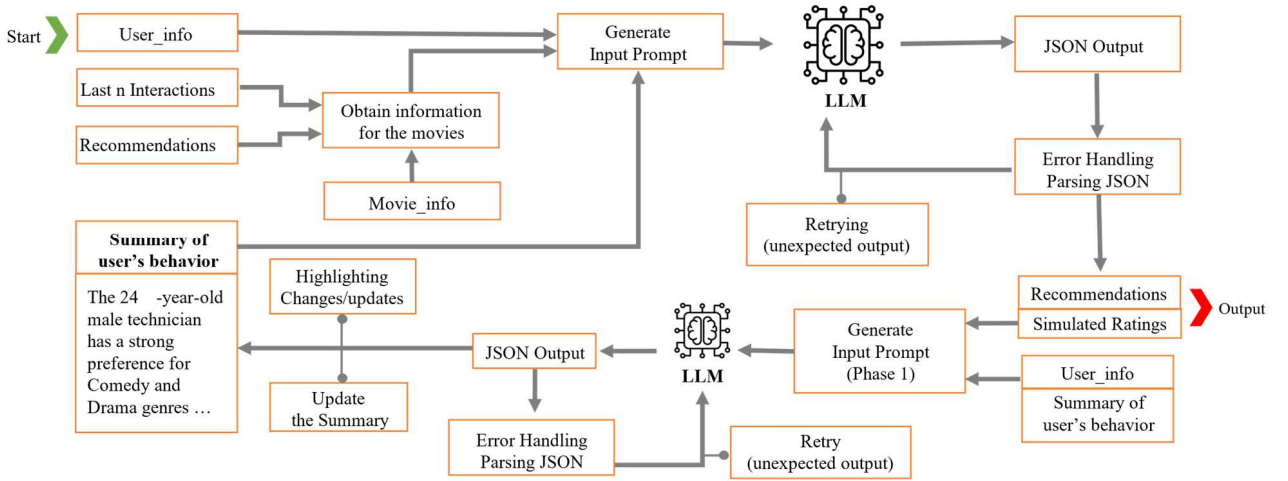

Fig. 3. Schematic view of Phase 2: generating simulated ratings and updating the profiles.

### III. PROMPT ENGINEERING

Prompt engineering, the practice of meticulously crafting input prompts to steer large language models (LLMs) outputs, has rapidly emerged as a pivotal aspect of AI interactions. This technique leverages the inherent capabilities of LLMs by framing queries and instructions in ways that elicit more accurate and relevant responses. Recent studies have delved into various strategies and methodologies for effective, prompt design, underscoring its significance in enhancing model performance across diverse applications [20], [21]. Prompt engineering was integral to our methodology, enabling effective interaction with the LLM at each step of the simulation process. We designed prompts for generating initial user summaries, updating these summaries, and producing simulated ratings. By providing clear and concise instructions, we ensured the LLM accurately simulated dynamic user feedback, demonstrating Lusifer's capability to emulate realistic changes in user behavior over time. This approach contributes to a scalable and adaptable framework for user simulation in online recommender systems. Sample prompts are available in the GitHub repository of the project[4].

We employed one-shot prompting techniques to achieve the desired output formats, including a single example in the prompt to guide the LLM's responses. Additionally, we implemented validation layers to process and verify the outputs generated by the LLM, both for simulated ratings and updating user profiles. In the rare cases where outputs could not be validated using our parsing protection layers, we introduced a retry mechanism that resends the prompt to the LLM a specified number of times. This approach effectively resolved most unexpected outputs and minimized system interruptions.

Furthermore, we incorporated error handling procedures to manage specific common issues, particularly those related to generating simulated ratings in JSON format. After each step, we stored the most recent user data, including the updated behavioral summary and simulated ratings. This allowed us to resume the process from the last successful point in case of unexpected interruptions, ensuring robustness and continuity in the simulation.

### IV. SCALABILITY AND ADAPTABILITY

Lusifer is designed as a scalable and adaptable environment for diverse applications in recommender systems. To utilize Lusifer, researchers can OpenAI API or any open-sourced LLM integrated with Ollama. Utilizing the OpenAI API allows us to leverage the advanced capabilities of GPT models without the overhead of computation needed for open sourced LLMs. On other hand, local LLM can bring scalability, speed, better error handling and customization without concerning rate limits. For this study, we used gpt-4o-mini from OpenAI, and Gemma:3B and Gemma:12B due to their efficient balanced of accuracy, speed and cost. Researchers can change the model within the configuration step to examine other APIs or Ollama supported models and provide comparative experiments.

The process begins with data preprocessing, where comprehensive textual features for users and items—such as user_info and item_info—are created to encapsulate all relevant information. This ensures that the LLM has sufficient context to simulate user behavior accurately.

In configuring the simulation environment, these textual features are provided to the LLM along with designed prompts, which can be adjusted based on specific research needs. Researchers can fine-tune parameters like the number of previous interactions to consider (e.g., the last 40 movies) and batch sizes for updating user summaries (e.g., batches of 10). This flexibility allows for optimization of the simulation's performance and computational efficiency.

Lusifer analyzes user behavior and generates updated summaries of user preferences, augmenting the original user data with an additional feature containing the user behavior summary while storing intermediate user-profiles and explanations of the changes. A test set is provided for experimentation, and batch sizes are specified to generate simulated ratings. The output includes the test set enriched with a simulated_ratings feature produced by the LLM.

When training reinforcement learning (RL) agents, recommendations from the agent are supplied as a data frame containing item IDs. Lusifer returns the simulated ratings and updates the user's behavior by incorporating new LLM-generated ratings. Batch sizes can be specified to control the frequency of updating the RL agent's behavior, enabling scalable and efficient training processes.

### V. EXPERIMENTAL RESULTS AND DISCUSSION

We limited the responses for Lusifer and all baselines to provide only integer ratings from scale 1 to 5 (no decimals allowed) to have the experiment focusing on simulating real case scenario, As of real-users ratings. Table 1 shows that the baseline methods achieved robust predictive performance, with lower RMSE, MAE and Pearson correlations. These metrics highlight the baselines' effectiveness in leveraging global patterns across the user base, even when restricted to each user's last 40 interactions. These models capitalize on well-established collaborative filtering strategies by analyzing collective user data and identifying shared behaviors. In contrast, Lusifer presents higher RMSE and MAE values along with a lower Pearson correlation, reflecting its narrower focus on a single user's recent history without drawing on the collective knowledge from other users. On the other hand, in cold start scenarios, where we have limited interactions from users (users with less than 10 previous interactions), Lusifer outperformed most baselines in all metrics using Gemma:12B open-sourced model. While Lusifer is not designed to maximize predictive accuracy, its core contribution lies in capturing individual users' evolving preferences. By simulating and updating user profiles at each interaction, Lusifer provides a more dynamic and responsive environment. This feature is particularly valuable for training reinforcement learning agents in adaptive recommender systems.

---
[4] Project Repository: https://github.com/danialebrat/Lusifer

TABLE I.    EXPERIMENTAL ACCURACY RESULTS

| Method | Movielens 100K | | | Movielens 1M | | |
|---|---|---|---|---|---|---|
| | *RMSE* | *MAE* | *Pearson* | *RMSE* | *MAE* | *Pearson* |
| SVD++ | 1.05 | 0.76 | 0.395 | 1.14 | 0.83 | 0.383 |
| NCF | 1.18 | 0.87 | 0.299 | 1.17 | 0.86 | 0.322 |
| RNN4Rec | 1.14 | 0.84 | 0.232 | 1.09 | 0.80 | 0.314 |
| ALS | 1.11 | 0.81 | 0.392 | 1.10 | 0.79 | 0.423 |
| 4o-Mini | 1.57 | 1.18 | 0.205 | 1.73 | 1.32 | 0.186 |
| Gemma3:4b | 1.39 | 1.02 | 0.259 | 1.50 | 1.10 | 0.190 |
| Gemma3:12b | 1.19 | 0.88 | 0.259 | N/A | N/A | N/A |

One of our key observations is that integrating unstructured textual metadata (e.g., movie overviews, tags, and user-based information) can significantly improve Lusifer's accuracy. LLMs excel at extracting nuanced information from these text fields, leading to more context-aware predictions and richer user profiles. Moreover, our experiments show that explicitly including numeric ratings in user profiles may sometimes reduce prediction accuracy, likely due to LLMs' more limited reasoning on numeric inputs. Limiting numeric exposure to only the most recent interactions (last 10 ratings) while relying primarily on textual metadata for user profiles helped Lusifer strike a balance between user history and new contextual clues, enhancing both the system's accuracy and its capacity for explainability.

TABLE II.    EXPERIMENTAL ACCURACY RESULTS IN COLD START SCENARIO (USERS WITH LESS THAN 10 INTERACTIONS IN MOVIELENS 100K)

| Method | Movielens 100K | | |
|---|---|---|---|
| | *RMSE* | *MAE* | *Pearson* |
| SVD++ | 1.11 | 0.83 | 0.366 |
| NCF | 1.29 | 0.99 | 0.224 |
| RNN4Rec | 1.19 | 0.9 | 0.221 |
| ALS | 1.35 | 1.04 | 0.297 |
| 4o-Mini | 1.51 | 1.14 | 0.226 |
| Gemma3:4b | 1.39 | 1.02 | 0.241 |
| Gemma3:12b | 1.18 | 0.88 | 0.222 |

Lusifer's user-centric design provides the ability to handle cold-start scenarios effectively as shown in Table 2. Traditional model-based methods often struggle when encountering a new user with minimal interaction data. Since Lusifer builds profiles from textual metadata (e.g., movie overviews, genre descriptions, or user attributes) and a small set of recent interactions, it can quickly adapt and simulate realistic feedback without extensive historical data. This also extends to cold-start items: Lusifer can still generate predictions by relying on the item's textual metadata even if a new item lacks prior ratings. This ability to manage both user- and item-level cold starts underscores Lusifer's versatility in real-world recommendation scenarios, where new content and new users constantly join the system.

In summary, while Lusifer does not achieve the same predictive precision as the baselines, it stands out as a flexible and explainable simulation environment. Its ability to handle cold-start situations, adapt to sparse data, and offer dynamic, step-by-step user updates makes it an ethically viable and practically useful tool for testing and training RL-based recommender systems under real-world constraints.

VI. CONCLUSION

This paper introduces Lusifer, an LLM-driven environment for generating dynamic user behaviors in RL-based recommender systems. By focusing on recent user interactions and iteratively updating profiles, Lusifer captures evolving preferences more realistically than static baselines, even under limited data. While Lusifer achieves a level of performance, comparable to traditional methods in terms standard accuracy metrics, its strengths lie in adaptability, scalability, explainability, and effective cold-start handling. These features position Lusifer as a scalable, ethical alternative to live user studies, paving the way for future research on RL-based recommendation strategies and enabling more profound insights into user preference evolution.

VII. FUTURE WORK

We aim to refine Lusifer's accuracy and expand its practical applications. First, we will enhance LLM precision by experimenting with advanced architectures, refining prompt engineering, and utilizing richer metadata to capture subtle user behaviors. Second, we plan to fully integrate Lusifer with RL-based recommender agents in continuous training loops, enabling direct assessments of adaptive strategies under shifting user profiles. Third, we will incorporate more diverse feedback signals—such as textual reviews, browsing behaviors, and session data—to enrich user modeling beyond explicit ratings. Fourth, we intend to explore multi-domain scenarios, extending Lusifer from movie recommendations to other fields like e-commerce, news, or music, ensuring broader applicability. Lastly, we will conduct extensive benchmarking against established user-simulation platforms to validate Lusifer's effectiveness further and potentially position it as a standard for LLM-based simulated feedback in recommender system research. Through these enhancements, Lusifer aims to offer a scalable, adaptable, and ethically robust environment for the next generation of RL-driven recommender systems.